\documentclass[oldversion]{aa} 
\usepackage{graphicx}
\usepackage{aalongtable}
\usepackage{txfonts}
\usepackage{rotating}
\usepackage{lscape}
\usepackage{natbib}
\newcommand{\gsim}{\;\lower.6ex\hbox{$\sim$}\kern-7.75pt\raise.65ex\hbox{$>$}\;}
\newcommand{\lsim}{\;\lower.6ex\hbox{$\sim$}\kern-7.75pt\raise.65ex\hbox{$<$}\;}
\newcommand{\teff}{$T_{\rm eff}$}

\newcommand\msun{M$_\odot$}

\begin{document}
\title{Searching for multiple stellar populations in the massive, old open cluster Berkeley~39\thanks{Based on observations collected at 
ESO telescopes under programme 386.B-0009}\fnmsep\thanks{
   Tables 2 and 3 are only available in electronic form at the CDS via anonymous
   ftp to {\tt cdsarc.u-strasbg.fr} (130.79.128.5) or via
   {\tt http://cdsweb.u-strasbg.fr/cgi-bin/qcat?J/A+A/???/???}}
 }

\author{
A. Bragaglia\inst{1},
R.G. Gratton\inst{2},
E. Carretta\inst{1},
V. D'Orazi\inst{3,4},
C. Sneden\inst{5}
\and
S. Lucatello\inst{2}
}

\authorrunning{A. Bragaglia et al.}
\titlerunning{The open cluster Be~39}

\offprints{A. Bragaglia, angela.bragaglia@oabo.inaf.it}

\institute{
INAF-Osservatorio Astronomico di Bologna, Via Ranzani 1, I-40127
 Bologna, Italy
\and
INAF-Osservatorio Astronomico di Padova, Vicolo dell'Osservatorio 5, I-35122
 Padova, Italy
I-35122 Padova, Italy
\and
Department of Physics and Astronomy, Macquarie University, Balaclava Rd,
North Ryde, NSW 2109, Australia
\and
Monash Centre for Astrophysics, School of Mathematical Sciences, Building
28, Monash University, VIC 3800, Australia
\and
Department of Astronomy and McDonald Observatory, The University of Texas, Austin, TX 78712, USA
  }

\date{}

\abstract{
The most massive star clusters include several generations of  stars with a
different chemical composition (mainly revealed by an Na-O anti-correlation)
while low-mass star clusters appear to be chemically homogeneous.   We are
investigating the chemical composition of several clusters with masses  of a few
$10^4~M_\odot$\ to establish the lower mass limit for the multiple  stellar
population phenomenon.  Using FLAMES@VLT spectra we determine abundances of Fe,
O, Na, and  several other elements ($\alpha$, Fe-peak, and neutron-capture
elements) in the  old open cluster Berkeley 39.  This is a massive open
cluster:  $M \sim 10^4~M_\odot$, approximately at  the border between small
globular clusters and large open clusters.   Our sample size of about  30 stars
is one of the largest studied for  abundances in any open cluster to date, and
will be useful to determine improved  cluster parameters, such as  age,
distance, and reddening when coupled  with precise, well-calibrated
photometry.   We find that Berkeley 39 is slightly metal-poor, $<$[Fe/H]$>$~=
$-$0.20,  in agreement with previous studies of this cluster.   More
importantly, we do not detect any star-to-star variation in the  abundances of
Fe, O, and Na within quite stringent upper limits.  The r.m.s. scatter is 0.04,
0.10, and 0.05 dex for Fe, O, and Na,  respectively. This small spread can be
entirely explained by the noise in the spectra  and by uncertainties in the
atmospheric parameters.  We conclude that Berkeley 39 is a  single-population
cluster. 
}


\maketitle

\section{Introduction} \label{intro}

\begin{figure}
\centering
\includegraphics[scale=0.44]{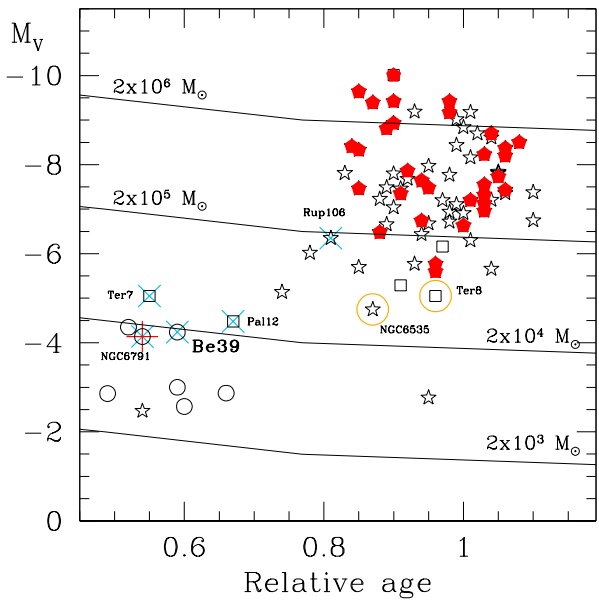}
\caption{Relative age parameter vs. absolute magnitude $M_V$\ for globular and
old open clusters.  Red filled symbols are GCs where the Na-O anti-correlation
has been  observed;  open star symbols and open squares mark MW and Sagittarius
GCs, respectively, for which not enough data are available; open  circles are
old open clusters (data from \citealt{lata}).  Finally, light blue crosses
indicate several clusters  that do not show evidence of Na-O anti-correlation:
two GC  members of Sagittarius dSph, one of the main body (Ter~7) and  one of
the stream (Pal~12); the new entry Rup~106; and two  massive, old OCs, Be~39 and
NGC~6791 (the latter indicated also  in red, since  its situation is still not
completely assessed). Two orange, large circles indicate the other GCs of our
project. Superimposed are lines of constant mass (light solid lines, see
\citealt{bellazzini}).}
\label{fig-mvage}
\end{figure}

Most, perhaps all, Galactic globular clusters (GCs) host multiple  stellar
populations that can be identified by variations of  Na and O abundances
\citep[e.g,][ and references therein]{zcorri}, which are anti-correlated with 
each other \citep[see e.g.,][for recent reviews]{araa,grattonrev}.  On the other
hand, Na-rich and O-poor stars that are common in GCs are rare  among field
stars (\citeauthor{zcorri}) and are possibly absent from open  clusters 
\citep[OCs, see e.g.,][]{desilva}.  A similar result is obtained for other light
elements, such as C and N, see e.g., \cite{martell1}, \cite{martell2}.  The
onset  of the  mechanism that leads to the multiple stellar population
phenomenon is tied to the dynamics of the star formation process and has in turn
to depend on many variables such as  age, metallicity, total mass, and probably
also on the environment (e.g., position in the Galaxy or in a smaller galaxy)
where the cluster formed.   According to \cite{zcorri},  however, there seems to
be a minimum (present-day) threshold mass of about a few $10^4~M_\odot$,
as shown in Fig.~\ref{fig-mvage} (an updated  version of the $M_V$-age plot
originally presented in \citealt{zcorri}).

Unfortunately, few clusters with masses close to this threshold value  have
been surveyed extensively enough to show whether they really have Na  and O
abundance variations \citep[see e.g.,][]{zcorri,bragagliaaas}. Two GCs in this
mass interval, Ter~7 and Pal~1 (Fig.~\ref{fig-mvage}) seem to lack these
variations; however, only a handful of stars in each of  them has been studied
\citep{tautvaisiene,sbordone1,sbordone2,cohen}.   Another case seems to be
Rup~106 \citep{rup106}, at a higher present-day mass.  A possible case for
variations in the abundances of these elements has recently  been made for the
massive old open cluster NGC~6791  \citep[][but see the conflicting
conclusion reached  by our group, Bragaglia et al, in preparation]{geisler}.

Observation of a larger sample of clusters is required, with the  same kind of
large samples we obtained for more than 20 GCs using the multi-object,
high-resolution spectrograph FLAMES@VLT  \citep[see
e.g.,][]{carretta09a,carretta09b}.  We have started systematic observations of
clusters in this transition  region, including both the most massive metal-rich
disc open clusters and the faint end of the mass distribution of metal-poor
globular clusters.  This approach may help to understand if the only factor at
play is mass  (remembering of course that $M_V$ measures the present-day  mass, 
not directly the original one), or if other factors, such as the environment 
in which the clusters formed, have to be considered.

In this paper we present results for the OC \object{Berkeley 39}, an old (about
6 Gyr, \citealt{kassis} and Sect.~\ref{obsphot}) massive OC ($M_V=-4.28$,
\citealt{lata}, indicating a mass of  $\sim2\times10^4~M_\odot$, see
Fig.~\ref{fig-mvage}). Be~39 is located in the outer part of the disc, at about
11 kpc from the Galactic centre \citep[see][]{friel10}.   Friel et al.
determined the chemical composition of this cluster from  high-resolution
spectra of four red giants.  These authors showed that this cluster is slightly
metal-poor  ([Fe/H]=$-0.21\pm 0.01$),  which agrees very well  with the
metallicities of other clusters at similar Galactocentric distances (see
their Fig.~3).  Friel et al. obtained very consistent results from their four
stars.  However, their sample is too small to exclude a significant spread  in
abundances in the cluster, as would be expected from the multiple  population
phenomenon.  Extant photometric data are not sufficiently accurate to shed light
on this question (see Sect.~\ref{obsphot}).  Chemical inhomogenieties in Be~39
cannot be ruled out at present. Therefore  we included Be~39 in our survey, and
obtained abundances from  high-resolution spectra for several tens of member
stars, a substantial  fraction of the evolved population of this cluster.  

While the spectral regions of our data were optimised to discuss the Na-O 
anti-correlation, our spectra have a wider interest. Open clusters are good
tracers of the properties of the disc of our Galaxy, see e.g., 
\cite{friel95,bt06,yong12} and references therein. Notwithstanding many
observational efforts, only a fraction of the known Galactic OCs \citep[about
2000, according to the catalogue in][and its updates]{dias} have been observed
with high-resolution spectroscopy. Searching the literature, we found only about
80 with an age older than 100 Myr (see Sect.~\ref{resu}) and in general only a
handful of stars were observed in each of them. If one aims to investigate the
history of the Galactic disc (formation and evolution), the old clusters are
very important; our observations provide further, improved information on one of
the few very old OCs. The large sample of stars is also very useful, because it
permits us to put stringent constraints to the internal abundance homogeneity,
something that is  relevant for the cluster formation mechanisms. 

The paper is organized as follows: in \S2 we present the basic data  for this
cluster, the selection of stars to be observed, and the observations
themselves.  In \S3 we describe the analysis method; a critical point is the 
derivation of accurate atmospheric parameters.  In \S4 we present and discuss
the results of our analysis.  \S5 contains the summary and conclusions.

\section{Observations}\label{obs}

\begin{figure*} 
\centering
\includegraphics[bb=30 190 570 480, clip,scale=0.8]{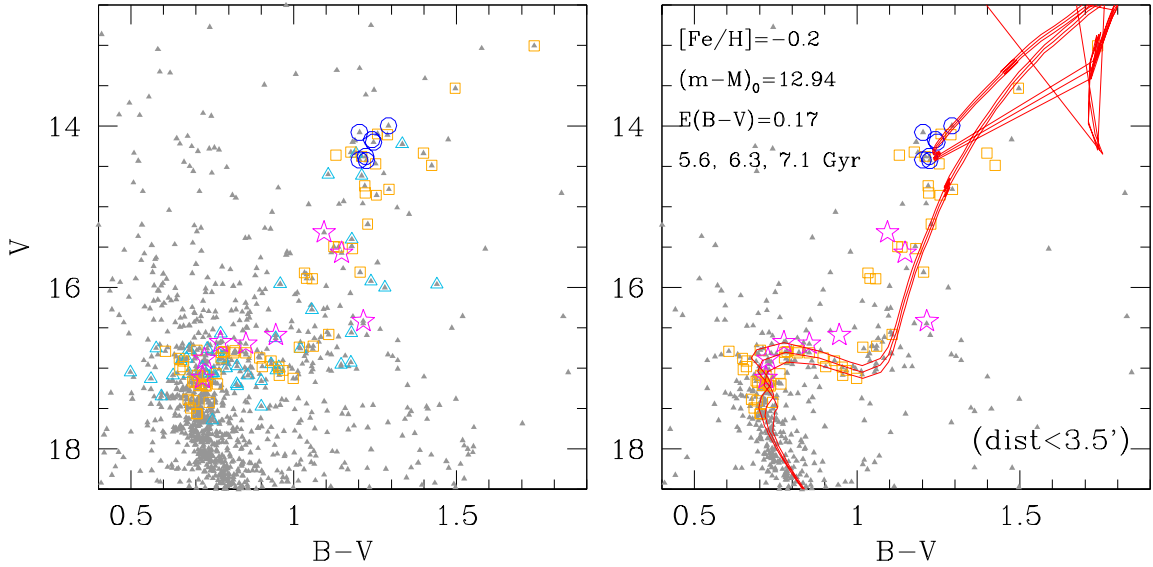}
\caption{Left: CMD for Be~39 \citep{kassis} with observed stars  indicated by
different symbols: UVES target (blue circles),  GIRAFFE targets (orange squares
for members, cyan triangles  non-members), binaries (magenta stars).  Right:
CMD for the inner 3.5 arcmin, with  observed member stars indicated by  larger
symbols and three isochrones shown.  The parameters used for the fit are
indicated.}
\label{fig-cmd}
\end{figure*}

\subsection{Photometry and cluster parameters}\label{obsphot}

Photometric observations for Be~39 have been used to study its  variable stars
and to determine the cluster parameters \citep[see e.g.,][]  {kr,kassis}; a
summary of past results can be found in \cite{friel10}. Of the publicly
available optical photometry, \cite{kr} collected $B,V$ data on a small field of
view (4.5$\times$7.2 arcmin$^2$) while \cite{kassis} provided $B,V,I$ photometry
on a 14.7$\times$14.7 arcmin$^2$ area, better suited to the FLAMES spectrograph
field \citep[25\arcmin \ in diameter,][]{pasquini}. We obtained the two
photometric catalogues through the database  WEBDA\footnote{The OC database is
at  {\em http://www.univie.ac.at/webda/webda.html}}    and used the second set
to select our targets (see Fig.~\ref{fig-cmd}). 

In our series of papers on the Na-O anti-correlation in GCs we used  atmospheric
parameters directly derived from the photometry, so we  checked if it was
possible to do the same for Be~39 as well.  Unfortunately, the two photometric
data sets are not identical; they show an average difference of $-$0.02
(rms=0.07) mag in $V$ and $+$0.06 (rms=0.10) mag in $B$ (in the sense Kassis et
al. minus Kaluzny \& Richtler)  based on about  730 stars in common. This could
seem a minor effect, but a difference of about 0.07-0.08 mag  in $B-V$ means a
large difference in stellar effective temperatures in Be~39. Furthermore, the
largest discrepancies between the two sets are found also among the brightest
stars, especially the red giant branch  (RGB) and red clump (RC) ones.  We were
unable to unequivocally establish which, if either, of the  two photometric sets
is the best calibrated, so we adopted the \cite{kassis} one because it is
available over a larger field of view. However, we manually adjusted a few
values for stars that are  radial-velocity (RV) members of the cluster but have
magnitudes and colours  that are discrepant from the evolutionary sequences (see
Sect.~\ref{param}) in one of the CMDs.

We also re-determined the cluster parameters, using the Padova isochrones 
\citep{marigo} retrieved from the interactive database {\em
http://stev.oapd.inaf.it/cgi-bin/cmd} at the metallicity derived by 
\cite{friel10}, i.e., [Fe/H]=-0.20.  Fig.~\ref{fig-cmd} (right panel) shows the
result; the age is between 5.5  and 7 Gyr, with a best-guess age of about 6
Gyr.  The distance modulus is $(m-M)_0=12.94$, and the reddening is 
$E(B-V)=0.17$, slightly higher than found in previous studies (but consistent, 
considering the relatively large uncertainties in the photometric data). These
values were used to determine effective temperatures and surface gravities of
the programme stars (Sect.~\ref{param}). Note however from the figure that RGB
stars do not define a narrow sequence;  this could be partly because of
undetected binaries, but could  also indicate differential reddening and/or
uncertain photometric values.

We could not locate Str\"omgren colours for Be~39, which could have been used to
test the uni- or multi-population nature of the cluster, as  discussed on
empirical grounds by \cite{carrettastro} and on theoretical  ones by
\cite{sbordonestro}. The latter paper also considered the more usually available
Johnson-Cousins bands; however, as we have noted, existing broadband photometry
in those filters is not sufficiently precise for this goal.

\subsection{FLAMES data}\label{obsspec}

We obtained six exposures of Be~39 with the multi-object  spectrograph
FLAMES@VLT, as in our previous studies on the Na-O  anti-correlations in
clusters. The observations were obtained in service mode and were spaced in
time  (see Tab.~\ref{tab-log}) to search for RV variations that would
indicate the presence of binary systems. We used the UVES 580nm setup
($\lambda\lambda\simeq4800-6800$~\AA) and  the GIRAFFE high-resolution grating
HR13, which contains the [O {\sc i}]  line at 6300 \AA \ and the Na {\sc i}
doublet at 6154-6160 \AA.

We selected Be~39 candidates using available information on RV from 
\cite{friel02} and \cite{friel10} to help selecting only probable cluster 
members.  All chosen targets are free from neighbours: they have no other
stars  closer than 3\arcsec \ (or 2\arcsec \ if they are at least 2 mag fainter 
than the desired candidates). As shown in Fig.~\ref{fig-cmd}, we selected stars
on the RC, or close to it,  for the UVES fibres; only one configuration was
prepared, so we have  seven stars observed at high-resolution ($R\simeq45000$),
with the eighth  fibre used for sky subtraction.  The GIRAFFE fibres (at
$R\simeq22500$) were  allocated to other RC and RGB  objects and to stars on the
subgiant branch (SGB) and the turn-off (TO)  of the main sequence; 16 fibres
were used for sky correction.  For the SGB and TO stars the spectral region is
not ideal, since the  [O {\sc i}] line is too weak to be measured.  The Na {\sc
i} features of these stars still could be used to search for a  possible Na
abundance spread. However, the main targets of our observations are RGB and RC
stars, for  which Na and O can be measured; the most interesting information
provided  by fainter stars is their RV (i.e., membership).  Information on all
observed stars can be found in Tab.~\ref{tab-phot}; we  give the ID in the
\cite{kassis} catalogue, the ID in WEBDA, equatorial  coordinates, $V$, $B-V$
and $V-I$ colours (with a flag indicating stars  corrected as described above),
and the heliocentric RV.

\begin{figure}
\centering
\includegraphics[bb=40 150 380 700,clip,width=9cm]{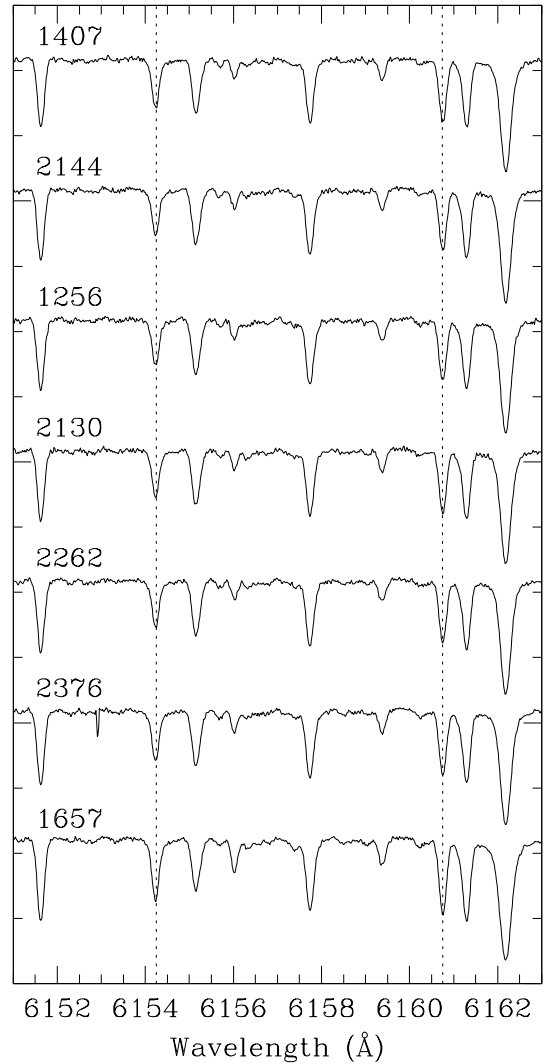}
\caption{Portion of the UVES spectra in the Na {\sc i} lines region;  the two
Na lines are indicated by the dotted lines. The normalised spectra are offset
for clarity and are plotted in  order of decreasing \teff, with the RGB star at
the bottom.  The Na and all other lines are practically identical in all stars, 
with only a small difference between the RGB and the RC ones.}
\label{fig-spena}
\end{figure}

The spectra were reduced (bias and flat field corrected, 1-D extracted, and
wavelength-calibrated) by the ESO personnel.  We applied sky subtraction and
division by an observed early-type star  (UVES) or a synthetic spectrum
(computed at the GIRAFFE spectral resolution) to correct for telluric features
near the [O {\sc i}] line, using the  IRAF\footnote{IRAF is distributed by the
National Optical Astronomical Observatory, which are operated by  the
Association of Universities for  Research in Astronomy, under contract with the 
National Science  Foundation.} routine {\em telluric}. The latter correction was
applied only to the UVES and bright GIRAFFE samples (RC, RGB stars), since this
[O {\sc i}] line is not  usable in the warmer faint GIRAFFE stars (SGB and TO).
We measured the individual RVs using {\em rvidlines} in IRAF and shifted  all
spectra to zero velocity before combining them.  The UVES final spectra have 
signal-to-noise ratio (S/N) in the range 100-170.  The S/N ratios (computed on small continuum regions) of
GIRAFFE spectra of RC stars are in the range 150-210;  of RGB  stars are 
generally between 100 and 200, and those of SGB and TO stars are 25-70. Values
for individual stars are listed in Tab.~\ref{tab-phot}.  A portion of the UVES
spectra showing the Na {\sc i} lines region is  shown in Fig.~\ref{fig-spena}.

\begin{figure}
\centering
\includegraphics[scale=0.44]{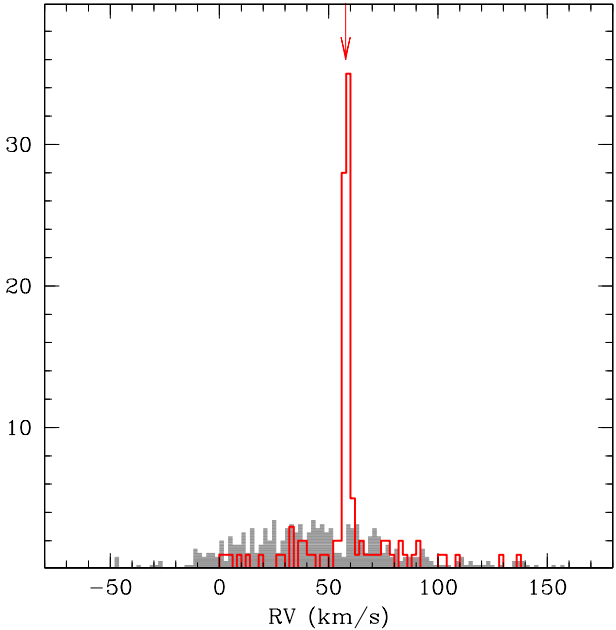}
\caption{Histogram of all RVs for the FLAMES spectra (red, open histogram)  and
for the Besan\c con model \citep{robin} in the direction of  Be~39, on the same
area of the  \cite{kassis} data and normalised  to the number of our stars
(filled, grey histogram). }
\label{fig-besa}
\end{figure}

The cluster average RV was computed separately for UVES and GIRAFFE spectra. We
found $<RV_{UVES}>=57.35$ (rms=0.85) km~s$^{-1}$ and  $<RV_{GIRAFFE}>=58.21$
(rms=1.62) km~s$^{-1}$.  We assume that all stars within $\pm3\sigma$ from the
average are cluster  members, with some more dubious cases indicated in
Tab.~\ref{tab-phot}.  Both these values compare well to RV=58.65 (rms=2.34)
km~s$^{-1}$  \citep{friel10} and  RV=55.0 (rms=2.9) km~s$^{-1}$
\citep{frinchaboy}.  Fig.~\ref{fig-besa} shows the RV  distribution of our stars
and compares it  to the expected distribution of field star velocities,
according to the  Besan{\c c}on model \citep{robin}.  Be~39 stands out
conspicuously and fewer than about ten field stars have  RVs similar to the cluster
one.

After pruning the sample using RVs we found a total of 67 possible single member
stars. All  seven UVES stars are RV members; of the GIRAFFE stars, 60 are
possible members, eight  are binary systems, and 43 are non-members.  This is
one of the largest sample of spectroscopically studied stars in OCs.

\begin{figure} 
\centering
\includegraphics[bb=50 164 580 540, clip,scale=0.44]{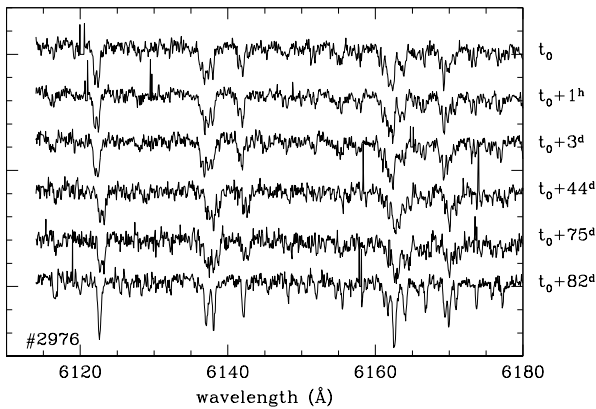}
\caption{Six spectra obtained for star \#2976, a binary near the  base of
the RGB.  The right axis indicates the time since the first observation.}
\label{fig-bin}
\end{figure}

While measuring the RVs we found a few stars with discrepant values at different
observation times, which is indicative of binary systems. On closer inspection,
some of them also displayed changes in spectrum  appearance; an example is shown
in Fig.~\ref{fig-bin}. There is also a suggestion, combining our RVs with those
of \cite{friel10}  (see Sect.~\ref{conf}), that one of the UVES and one of the
GIRAFFE  targets may be binary. If all these systems are members, this indicates
a binary fraction of  10/75  or 13\%. This is only  a lower limit and it
perfectly conforms to the fraction of binaries usually found in OCs from photometry \citep[see e.g.,][]{bt06}.

\begin{table}
\centering
\caption{Log of the observations}
\begin{tabular}{cccccc}
\hline\hline
OB  & UT date & UT$_{init}$ & exptime & airmass &seeing\\
       & (Y-M-D) & (h:m:s)      & (s)           &                &(arcsec)\\
\hline
A & 2010-12-10 & 06:26:33.886 & 2345& 1.085 &0.51 \\
B & 2010-12-10 & 07:19:18.192 & 2450& 1.064 &0.48 \\
C & 2010-12-13 & 05:07:19.888 & 2450& 1.205 &0.66 \\	
D & 2011-01-26 & 04:14:19.507 & 2450& 1.064 &0.55 \\
E & 2011-02-27 & 02:59:58.770 & 2450& 1.096 &1.26 \\
F & 2011-03-06 & 02:35:28.955 & 2345& 1.100 &1.30 \\
\hline
\end{tabular}
\label{tab-log}
\end{table}

\section{Analysis}\label{analysis}

\subsection{Atmospheric parameters}\label{param}

For this cluster we cannot apply the same procedure as we adopted for the GC
survey \citep{carretta09a,carretta09b}.  First, we saw that the photometry alone
cannot guarantee atmospheric  parameters with the required precision (at least,
not for all stars).  Second, we cannot use the IR K magnitude as we have
previously done  for other clusters to  determine the temperature from the $V-K$
colour.  Our sample does  not only include bright stars for which the  2MASS
photometry would be precise enough, but also faint stars for which  it would be
unreliable. We could potentially infer false inhomogeneities were we to use 
this colour. On the other hand, $B,V$, and $I$ photometry accurate enough (apart
from the mismatches discussed in Sec.~\ref{obsphot}) is available for all
targets.  We therefore opted for temperatures derived from $B-V$ and $V-I$ as 
input values.

We used the reddening and distance modulus derived above, adopting $E(V-I)=1.3
E(B-V)$, $A_V=3.1 E(B-V)$, and 4.75 as bolometric magnitude of the Sun.  We used
the colour-temperature relations of \cite{rm05} to find \teff, and averaged the
values obtained from $B-V$ and $V-I$.  The gravity was derived as in previous
papers following the relations of  \cite{alonso} for the bolometric correction
and a mass of 1.15 \msun,  as deduced from the isochrones.  Average photometric
\teff \ and $\log g$ values were computed,  weighting more those obtained from
the original colours than the   ``corrected" ones (see Sect.~\ref{obsphot}).

These \teff's and $\log g$'s were used to derive first-pass iron abundances  for
all stars in our sample.  The microturbulent velocities $v_t$ were obtained from
the relation  $v_t=-0.322~\log{g}+2.22$\ (\citealt{gratton96}).  Adopting this
relation, only very weak trends in the relation between  abundances from Fe
neutral lines and expected line strength  \citep[see][]{magain} were apparent,
with negligible effects on the abundances. We then computed the \teff \ and
$\log g$ values required to satisfy the  excitation and ionisation equilibria
(we recall, however, that there are only  three useful Fe {\sc ii} lines in the
GIRAFFE HR13 wavelength range).  The final adopted parameters are the average of
the spectroscopic and  photometric ones.  Metallicities for all stars were then
obtained with model  atmospheres interpolated in the \cite{kur} grid of model
atmospheres  (without convective overshooting) with atmospheric parameters 
whose abundance matches that derived from Fe {\sc i} lines.

\begin{figure*}
\centering
\includegraphics[scale=0.8]{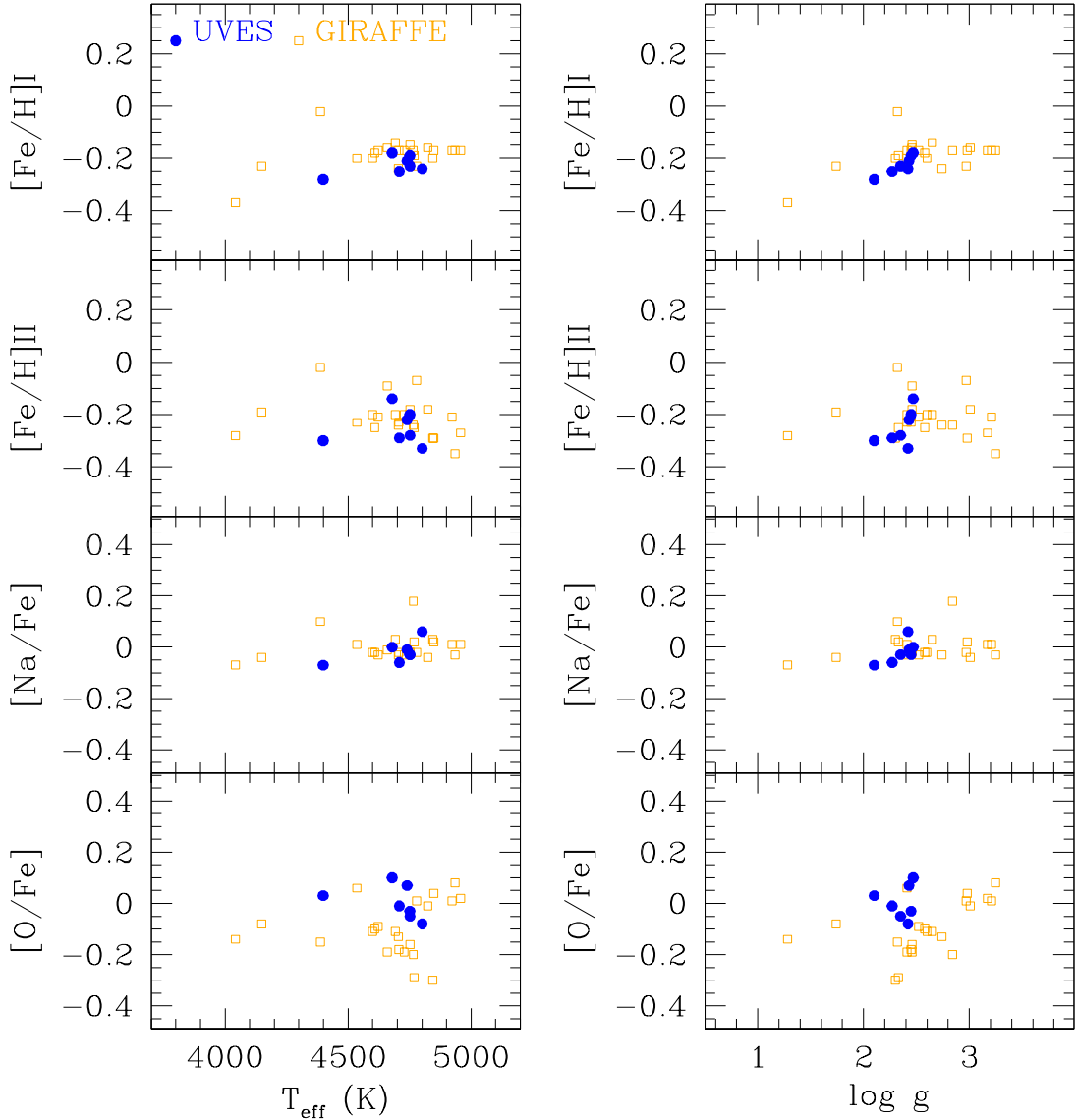}
\caption{Run of Fe {\sc i}, Fe {\sc ii}, Na, and O with \teff \ and  $\log g$
for stars on the RGB and RC.  Filled blue symbols are for UVES, open orange
squares for GIRAFFE targets.}
\label{fig-trend}
\end{figure*}

Temperature, gravity, microturbulent velocity, and Fe {\sc i} and {\sc ii}  are
given in Tab.~\ref{tab-abu1} for the UVES and Giraffe bright sample  (RGB and RC
stars).  The run of iron abundances with \teff \ and $\log{g}$ are shown in the
upper panels of  Fig.~\ref{fig-trend}.  These abundances do not display any
significant trends with \teff\ (left-hand panels) or $\log{g}$ (right-hand
panels).  The weak trend that is possibly presented by the lower gravity stars
in the [Fe/H] {\sc i}-$\log{g}$ plane is negligible in the context of the
present paper and could be caused by a variety of sources, such as blends,
continuum tracing, or inadequacy of the model atmospheres - all more noticeable
at cooler temperatures. We note, however, that there seems to be a small offset
between the  mean iron abundances of UVES and GIRAFFE stars. This is most
probably due to the combination of higher resolution  (hence accuracy in
equivalent width, EW, measurement, see next section) and higher S/N for the UVES
spectra. It is not worrisome, but in the next sections we will keep separate 
abundances from the two samples.

The values for the SGB and TO stars have larger errors and larger scatter
because of the low S/N combined with the warm temperatures and the larger errors
in the photometry. Furthermore, we could not derive O abundances for the faint
sample: the stars  are too warm to measure the [O {\sc i}] line, but not warm
enough for the  triplet at 6155-6158 \AA. While we proceeded with a preliminary
analysis also for the faint sample, we eventually  decided to discard these
stars and keep only the bright sample, for which our conclusions are sounder.  

\subsection{Equivalent widths, sensitivity, and errors} \label{sens}

We measured EWs with the software package Rosa \citep{rosa},  as described in
\cite{bragaglia01}. We adopted the same automatic procedure to define  the
local  continuum around each line as in previous papers, but we did not correct
the  GIRAFFE EWs to the UVES system because we did not have any star in common. 
Small differences between the two sets of EWs related to the different
resolution of the spectra may indeed exist, and explain the small  differences
in the abundances.  The EWs of the member stars are made available only through
the CDS.

To evaluate the sensitivity of the derived abundances to the adopted 
atmospheric parameters we repeated our abundance analysis by changing only  one
parameter each time. The amount of the variations in the atmospheric parameters
and the resulting  response in abundance changes of Fe, O, Na, and all elements
measured   (i.e., the sensitivities) are shown in Tab.~\ref{tab-sens}.  In the
upper part of the same table we also give the error contributed  by each
parameter.  The column labelled ``total error" gives the total error expected
from uncertainties in the atmospheric parameters and in the EWs, to be compared
to the observed scatter in order to understand if we have a statistically
significant variation in any of the elements.  Note that for some elements
(including Ca, V, and Ba) the observed  scatter is larger than the observational
errors.  However, for various reasons errors may have been underestimated for
these elements: lines of Ca and Ba are strong and saturated, and those of  V are
affected by strong HFS effects.

Errors in our measures are the following.

\paragraph{\teff:} The errors in \teff \  were derived by examining the rms of
residuals along a quadratic $\log g-T_{\rm eff}$ relation for RGB stars (after
eliminating RC stars).  This residual is 29 K, so we assume an error of $\pm30$
K. \paragraph{$\log g$:} The error in $\log g$ is dominated by the errors in 
\teff \  (30 K, yielding an error of 0.006 dex) and in photometry (given  all
the uncertainties noted in Sec.\ref{obsphot},  we assumed 0.05 mag, yielding an
error of 0.02 dex).  This yields a total error of $\pm0.03$ dex.
\paragraph{[A/H]:} The error is not larger than the observed star-to-star
scatter, i.e., $\pm0.04$ dex. \paragraph{$v_t$:} This is the dominant source of
errors for iron. It was estimated by considering the star-to-star scatter of the
trends of  abundances with expected line strength: the error is 0.09 km/s.  We
assumed an error of $\pm0.10$ km/s.

\begin{table}
\centering
\setcounter{table}{3}
\caption{Sensitivity to errors}
\setlength{\tabcolsep}{1mm}
\begin{tabular}{lrrrrrrr}
\hline\hline
Element    &\teff  &$\log g$  &[A/H]  &$v_t$  &EW  &Total   &Observed \\
           &(K) & (dex) & (dex) & km~s$^{-1}$&                            & error  & rms\\
\hline
&\multicolumn{4}{c}{$\Delta$ parameter} &&&\\
&100   &0.30   &0.10	&0.2 &&&\\	  
&\multicolumn{4}{c}{Error} &&&\\
& 30   &0.03   &0.04	&0.1 &&&\\
\hline
Fe {\sc i}   &   0.071 &  0.014 &  0.011 & -0.093 &  0.013  & 0.053& 0.049\\ 
Fe {\sc ii}  &  -0.164 &  0.046 &  0.026 &  0.043 &  0.036  & 0.066& 0.067\\ 
O {\sc i}    &  -0.058 &  0.114 &  0.027 &  0.087 &  0.085  & 0.098& 0.119\\ 
Na {\sc i}   &   0.004 & -0.025 & -0.015 &  0.059 &  0.052  & 0.060& 0.050\\ 
Mg {\sc i}   &  -0.032 & -0.008 & -0.007 &  0.071 &  0.053  & 0.065& 0.066\\ 
Al {\sc i}   &  -0.004 & -0.019 & -0.013 &  0.069 &  0.054  & 0.064& 0.025\\ 
Si {\sc i}   &  -0.101 &  0.033 &  0.010 &  0.061 &  0.047  & 0.064& 0.065\\ 
Ca {\sc i}   &   0.035 & -0.063 & -0.014 & -0.008 &  0.019  & 0.024& 0.093\\ 
Sc {\sc ii}  &  -0.087 &  0.118 &  0.025 &  0.028 &  0.053  & 0.063& 0.060\\ 
Ti {\sc i}   &   0.079 & -0.020 & -0.021 &  0.003 &  0.036  & 0.044& 0.053\\ 
Ti {\sc ii}  &  -0.103 &  0.120 &  0.023 &  0.074 &  0.106  & 0.117& 0.052\\ 
V {\sc  i}   &   0.096 & -0.012 & -0.019 &  0.030 &  0.023  & 0.041& 0.129\\ 
Cr {\sc i}   &   0.091 & -0.008 & -0.024 &  0.011 &  0.087  & 0.092& 0.150\\ 
Mn {\sc i}   &   0.031 & -0.052 & -0.001 & -0.039 &  0.044  & 0.050& 0.021\\ 
Co {\sc i}   &  -0.027 &  0.036 &  0.008 &  0.062 &  0.090  & 0.096&      \\ 
Ni {\sc i}   &  -0.037 &  0.040 &  0.007 &  0.033 &  0.019  & 0.028& 0.045\\ 
Ba {\sc ii}  & -0.046  &  0.063 &  0.033 & -0.084 &  0.046  & 0.066& 0.120\\ 
\hline
\end{tabular}
\label{tab-sens}
\end{table}

\section{Results}\label{resu}

Abundances of all elements were derived from EWs (and from synthetic spectra 
for O).  Hyperfine splitting corrections were applied for Sc, V, and Mn.  We
used the same lines and line parameters as in the papers of the FLAMES  GC
survey \citep[e.g.,][]{carretta09b}, originally from \cite{gratton03}. Reference
solar abundances are explicitly given in Tab.~\ref{tab-ave}.

The average values for all elements measured are presented in
Tab.~\ref{tab-ave}, separately for the UVES and GIRAFFE samples, given the
diversity of resolution, wavelength coverage, and number of lines available in
the two cases.  For UVES the average was derived from seven stars; for GIRAFFE
it is typically derived from 21 stars.  As stated above, only the brighter
sample was considered.  Furthermore, star 11 was always excluded because is more
that $3\sigma$ away   from the cluster average velocity and its approximately
solar metallicity  is clearly higher than that of the other stars of Be~39.

\begin{figure}
\centering
\includegraphics[bb=30 440 380 695,clip,scale=0.7]{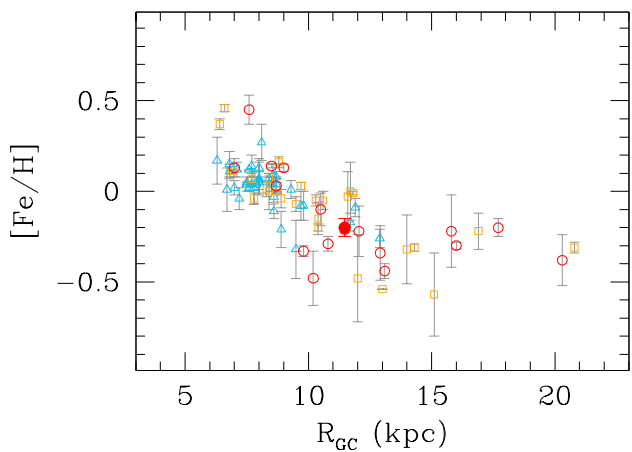}
\caption{Distribution of [Fe/H] with Galactocentric distance for OCs,
colour-coded according to age (light blue triangles : age 0.1-1 Gyr; orange
squares: age 1-4 Gyr; red circles: age older than 4 Gyr).  [Fe/H] and errorbars
are taken from the original literature sources. Be~39 is indicated with a filled
red symbol.}
\label{fig-grad}
\end{figure}

Be~39 clearly is a homogeneous cluster: the star-to-star scatter is always
explained by the errors expected from uncertainties in atmospheric parameters
and EWs (see Tabs.~\ref{tab-sens} and \ref{tab-ave}).  In particular, we find a
mean metallicity [Fe/H]=$-$0.23 (rms=0.04)  from UVES and  $-$0.18 (rms=0.06)
from GIRAFFE spectra. This agrees very well with \cite{friel10} (see also
Sect.~\ref{conf}). It is also in accord with metallicity expectations for
clusters at Be~39's  Galactocentric distance; the cluster lies at an R$_{\rm
GC}$ of about 11 kpc, where OCs generally have sub-solar metallicity, as shown
in Fig.~\ref{fig-grad}. Interestingly, its R$_{\rm GC}$ is around the transition
between inner and outer disc, where the metallicity gradient begins to flatten.
This is a region where more clusters should be observed to better constrain the
models of chemical evolution \citep[e.g.,][]{magrini09,friel10}. For the figure
we selected only clusters older than 100 Myr and with metallicity derived by
high-resolution spectra; the sample contains 79 old OCs. The picture is quite
clear; we caution, however, that the parameters (age, distance, metallicity)
have not been put on a homogeneous scale since this would be a great effort,
completely outside the goal of the present paper (see e.g., \citealt{yong12} for
a recent work on the abundances and trends in the outer disc.) We also confirm
Friel et al.'s derivation of a mild overabundance  of $\alpha$-elements
([$\alpha/{\rm Fe}]\simeq 0.1$.)

Elements Al, Ti {\sc ii}, and Mn could only be measured in the UVES spectra. It
is unfortunate that we could not measure Al also for the GIRAFFE sample,  since
it is involved in the light elements (anti-)correlations and may show  also very
strong star-to-star variations  \citep[see, e.g.,][for the GC
NGC~6752]{carrettaal}.  But the UVES results suggest that this is not the case
for  Be~39 (see Tab.~\ref{tab-ave}).

Barium is the only neutron-capture element we measured.   Its abundance was
derived from three lines in the UVES spectra at  5853.69, 6141.75,  and
6496.91 \AA.  These lines yield abundances that agree well, with a typical rms
of  0.12 dex for any given star. The star-to-star variations (0.08 dex) are due
to the uncertainties  in atmospheric parameters, especially in \teff \ and
$v_t$.   We have only a single Ba line (at 6141.75 \AA) in the GIRAFFE spectra. 
There is an offset between the Ba abundances measured in the two types of
spectra, which can be explained by differences in the EWs,  however.  The EWs of
the 6141 \AA\ are larger for GIRAFFE than for UVES stars of  similar atmospheric
parameters, most probably because of an unresolved  blend with Ni {\sc i} and Si
{\sc i} in the lower resolution spectra.  The contaminant-line contributions to
the GIRAFFE Ba lines seems confirmed  by estimates of their fractions  to the
total EW of the blended  Ba {\sc ii} feature.

\begin{figure}
\centering
\includegraphics[bb=30 450 330 680, clip,scale=0.8]{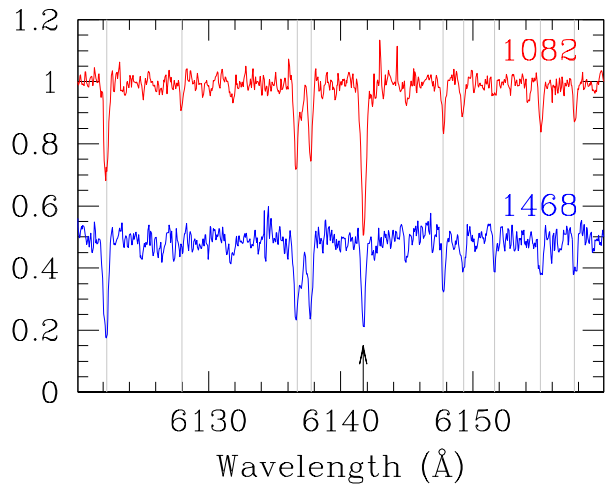}
\caption{Comparison of the Ba {\sc ii} 6141.7 \AA \ line (indicated  by an
arrow) in the Ba-star 1082 and in the star 1468, of very  close atmospheric
parameters, as confirmed by the similarity   of the other lines seen here (Ca,
Ni, two Fe {\sc i}, Fe {\sc i}+Fe {\sc ii}, Fe {\sc ii}, Fe {\sc i}, Na, and Si,
from left to right and indicated by grey lines). The spectra are shifted for
clarity.}
\label{fig-ba}
\end{figure}

Given the strength of the Ba {\sc ii} lines, which are always saturated,  Ba is
a difficult element to measure.  However, three stars of the faint sample (1082,
1357, and 2170) show a very  high Ba abundance;  all three have RVs compatible
with cluster membership. The first is the most convincing case suggestive of an 
s-process-enriched ``Barium star''.   Not only is its Ba line much stronger than
in stars of similar atmospheric  parameters and normal Ba abundance (see
Fig.~\ref{fig-ba}),  it  also shows a clear La {\sc ii} line at 6320.4 \AA, not
detected in other,  similar objects. Star 1082 is brighter than expected from a
single star at the TO and could be a  blue straggler; the high Ba could be due
to mass transfer in a binary system.  Since stars on the main sequence have only
a tiny convective envelope,  it does not need to have accreted a lot of mass to
explain the higher  Ba abundance; this would be possible even in a wide system
and this would  explain why its RV is not discrepant from the cluster' average
value.

\subsection{Na and O abundances}\label{nao}

The main goal of our study was determining the Na and O abundances  in Be~39
members to see whether they show an intrinsic star-to-star scatter  or even an
anti-correlation, the main spectroscopic signature of  multiple populations in
clusters.

We measured oxygen abundances from the [O {\sc i}] line using spectrum
synthesis.  We considered  the Ni {\sc i} contamination of the line  \citep[see
e.g.,][]{carretta07}, as well as the coupling of O and C  expected in cool
stars. The dissociation equilibrium was computed assuming [C/Fe]=-0.2 and
[N/Fe]=0.5, which are typical values for red giants in old OCs
\citep[see][]{Mikolaitis12}.  To estimate the sensitivity of our O abundances to
this assumption, we note  that they would be higher by $\sim0.05$~dex for RC
stars and by 0.11~dex  for stars at the tip of the RGB had we instead assumed
[C/Fe]=0. 

Only the Na doublet at 6154-6160 \AA \ was used to measure sodium  abundances
since the Na D and 5686-5690 lines appear only in the UVES  spectra and are too
strong at this metallicity. These abundances were corrected for NLTE effects
using the relation by  \cite{lind}\footnote{We generally use the
\cite{gratton99} corrections  in our works on GCs; however, given the much
higher metallicity of Be~39, we preferred to adopt the \cite{lind} method.  This
does not mean that we need to revise all our Na abundances, since  the older
NLTE corrections were computed for the metallicity regime of GCs.}.  We
interpolated their corrections with the quadratic relation  $A_{NLTE}-A_{LTE} =
-0.088 + 0.00076394 \times EW -0.0000144444  \times EW^2$, valid for EW$<150$
m\AA \ (that is, all Na lines in our stars).  The [Na/Fe] ratios (see 
Tab.~\ref{tab-abu1}) were then computed assuming [Fe/H]=-0.23 for UVES stars and
-0.18 for GIRAFFE stars.

\begin{figure}
\includegraphics[scale=0.44]{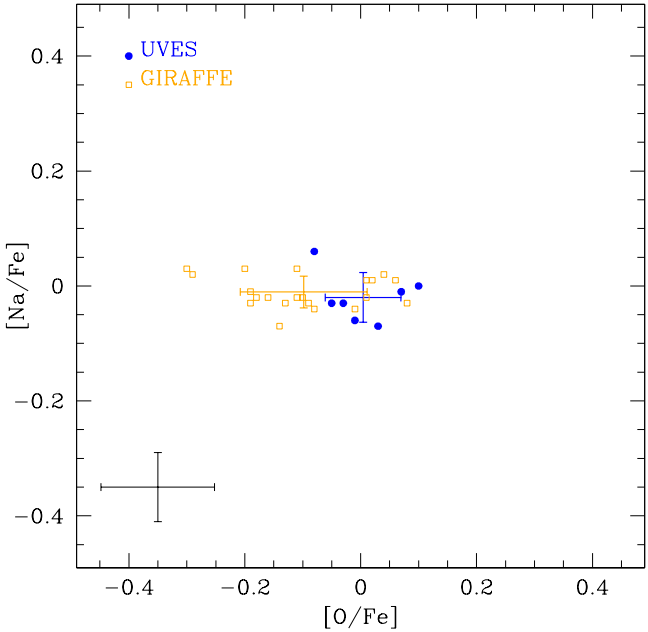}
\caption{Na and O distribution for the UVES (filled blue points) and the bright
GIRAFFE (open orange squares).  The rms in O and Na are shown by errorbars of
different colour.   The expected errors on O and Na abundances (see
Tab.~\ref{tab-sens})  are shown as black errorbars in the left lower corner of
the plot; they are comparable to the observed rms.}
\label{fig-nao}
\end{figure}

Neither O nor Na abundances show any significant trend with  \teff \ or $\log g$
(see lower panels of Fig.~\ref{fig-trend}). As apparent from Fig.~\ref{fig-nao}
and from the very small scatters  associated both to the O and Na values
(Tab.~\ref{tab-ave}),  Be~39 is very uniform in these elements as well.  The
star-to-star scatter is completely explained by the associated internal errors;
the only effect mimicking a relation between O and Na is the  small offset
between the UVES and GIRAFFE samples.  Even with all the caveats already
presented, this is a solid result:  Be~39 is a normal OC, without any signature
of the Na-O anti-correlation,  which is the typical ``tag" for a genuine GC
\citep{zcorri}.

\begin{table}
\centering
\caption{Average abundances for Be~39.}
\begin{tabular}{lcrrrrl}
\hline\hline
element     & Sun &abu &rms   &abu &rms & Notes\\
 & &\multicolumn{2}{c}{UVES} &\multicolumn{2}{c}{Giraffe} &\\
\hline
${\rm [Fe/H]}$ {\sc i}   & 7.54 & -0.23 &0.04  &-0.18 &0.06&\\
${\rm [Fe/H]}$ {\sc ii}  & 7.49 &-0.19 &0.05  &-0.21 &0.07&\\
${\rm [O/Fe]}$ {\sc i}   & 8.76 & 0.03 &0.07  &-0.08 &0.11&\\
${\rm [Na/Fe]}$ {\sc i} & 6.30  &-0.02 &0.04  &-0.01 &0.03& NLTE\\
${\rm [Mg/Fe]}$ {\sc i} & 7.43  & 0.17 &0.02  & 0.24 &0.05&\\
${\rm [Al/Fe]}$ {\sc i}  & 6.40 & 0.02 &0.03  &	&    &\\
${\rm [Si/Fe]}$ {\sc i}  & 7.53 & 0.12 &0.03  & 0.14 &0.05&\\
${\rm [Ca/Fe]}$ {\sc i} & 6.27  & 0.07 &0.05  & 0.17 &0.08&\\
${\rm [Sc/Fe]}$ {\sc ii} & 3.13 & 0.00 &0.05  &-0.03 &0.06&HFS\\
${\rm [Ti/Fe]}$ {\sc i}  & 5.00 & 0.03 &0.03  & 0.01 &0.06&\\
${\rm [Ti/Fe]}$ {\sc ii} & 5.07 & 0.12 &0.05  &	&    &\\
${\rm [V/Fe]}$ {\sc i}   & 3.97 &-0.08 &0.04  &-0.16 &0.08&HFS\\
${\rm [Cr/Fe]}$ {\sc i}  & 5.67 &-0.15 &0.05  &-0.13 &0.08&\\
${\rm [Mn/Fe]}$ {\sc i}  & 5.34 & 0.00 &0.02  &	&    &HFS\\
${\rm [Ni/Fe]}$ {\sc i}  & 6.28 & 0.04 &0.04  & 0.07 &0.03&\\
${\rm [Ba/Fe]}$ {\sc ii} & 2.18 & 0.14 &0.08  & 0.30 &0.09&\\  
\hline
\end{tabular}
\label{tab-ave}
\end{table}

\subsection{Comparison with \cite{friel10}} \label{conf}

\cite{friel10} observed two RC and two bright RGB stars using the echelle 
spectrograph on the  KPNO 4m telescope.  We give the data for the three objects
in common with our study in   Tab.~\ref{tab-conf}; two of them were observed
with UVES fibres, one with GIRAFFE.  The resolution of their spectra is
$R\sim28000$, intermediate between our  GIRAFFE and  UVES spectra, the
wavelength coverage is  $\lambda\lambda=4800-8100$ \AA, and the S/N  is always
high,  between 70 and 115. Friel and coworkers noted that one star (WEBDA ID
2619, or 2130 for \citealt{kassis},  adopted as our ID) is probably a binary,
combining their data and the  RV measured by \cite{frinchaboy}, differing by
about 10 km~s$^{-1}$.  We do not  see any strong indication of an RV change in
our data, but  the average RV is about 1.5 km~s$^{-1}$ lower than theirs.  A
similar behaviour is shown by star 1435/1139, again with our RV lower  by about
6 km~s$^{-1}$ but no strong signature of RV variation (only  one discrepant
value in our six measures).  We therefore define those two stars as probable
binaries.

Tab.~\ref{tab-conf} displays a direct comparison of the  \cite{friel10}
atmospheric parameters and ours; they generally agree well.  We also
compared our EWs with theirs; the agreement is excellent, with  our EWs slightly
smaller.  We find ${\rm EW_{UVES} = 0.992 (\pm0.019) \times EW_{Friel} -3.86}$
m\AA \  (with rms=0.64 m\AA, over 102 lines) and  ${\rm EW_{GIRAFFE} = 1.030
(\pm0.046) \times EW_{Friel} -1.1}$ m\AA \ (with rms=0.76 m\AA, over 26 lines).

A detailed comparison of the two analyses is beyond the scope of our paper.  We
only note that the derived abundances agree well (see Tab.~\ref{tab-conf}),
especially for iron and the $\alpha$-elements.  We also recall that we corrected
the Na abundances for NLTE, while Friel and collaborators did not, since these
corrections are small for such high-metallicity, warm stars.  Finally, we do not
provide here any comparison with the properties of other clusters, or discuss
the radial abundance distribution.   This is outside the main goal of our study;
moreover, the abundances  are so similar to those derived by \cite{friel10} that
all considerations and conclusions of these authors apply. 

\begin{table}
\centering
\caption{Comparison with \cite{friel10} for two UVES stars and one GIRAFFE star.}
\setlength{\tabcolsep}{1.15mm}
\begin{tabular}{ccccc ccccc c}
\hline\hline
Star  &\teff &$\log g$ & $v_t$  & RV  &  ID  &  \teff & $\log g$ & $v_t$ &  RV & U/G\\
\multicolumn{5}{c}{Friel et al. 2010} &\multicolumn{5}{c}{This paper} &\\
\hline
2055 &4450 & 1.8 & 1.5 &  57.5 &1657  & 4399 & 2.10 & 1.54 &  57.5 & U\\
2619 &4750 & 2.2 & 1.5 &  60.0 &2130  & 4740 & 2.43 & 1.44 &  58.5 & U\\
1435 &4750 & 2.2 & 1.5 &  62.8 &1139  & 4659 & 2.46 & 1.43 &  56.9 & G\\
\hline
\end{tabular}

\vspace*{5mm}
\setlength{\tabcolsep}{2mm}
\begin{tabular}{lrrrrr}
\hline
Element    &abu   &rms  &abu  &rms   &Friel+10 \\         
&\multicolumn{2}{c}{UVES} &\multicolumn{2}{c}{Giraffe} & \\
\hline 
${\rm [Fe/H]}$ {\sc i}   &-0.23 &0.04  &-0.18 &0.06	&   -0.21\\
${\rm [O/Fe]}$ {\sc i}   & 0.03 &0.07  &-0.12 &0.11	&    0.02\\
${\rm [Na/Fe]}$ {\sc i}  &-0.02 &0.04  & -0.01&0.03	&    0.09\\
${\rm [Mg/Fe]}$ {\sc i}  & 0.20 &0.02  & 0.22 &0.05	&    0.19\\
${\rm [Al/Fe]}$ {\sc i}  & 0.05 &0.03  &      &	        &    0.20\\
${\rm [Si/Fe]}$ {\sc i}  & 0.14 &0.05  & 0.14 &0.05	&    0.20\\
${\rm [Ca/Fe]}$ {\sc i}  & 0.10 &0.05  & 0.15 &0.08	&    0.01\\
${\rm [Ti/Fe]}$ {\sc i}  & 0.06 &0.03  &-0.01 &0.06	&    0.03\\
${\rm [Ti/Fe]}$ {\sc ii} & 0.15 &0.05  &      &	        &   -0.09\\
${\rm [Cr/Fe]}$ {\sc i}  &-0.13 &0.05  &-0.13 &0.08	&    0.06\\
${\rm [Ni/Fe]}$ {\sc i}  & 0.07 &0.04  & 0.05 &0.03	&   -0.06\\
\hline
\end{tabular}
\label{tab-conf}
\end{table}

\section{Summary and conclusions}\label{end}

We presented the abundance analysis of the high  mass, old OC
Be~39.  The high-resolution spectra of more than 100 stars, obtained with
FLAMES@VLT,  permitted us to isolate about 70 cluster members.    With the advent
of surveys targeting large numbers of stars in large and significant sets of
clusters (e.g.,  the Gaia-ESO Survey, recently started on FLAMES@VLT, see
\citealt{ges}), comparable samples will become more common, but this is
presently  the largest  sample of stars in this particular cluster and one of
the largest in general. The membership  information was used  to determine
updated cluster parameters (age, reddening, and distance).  About one half  of
the stars were analysed to determine metallicity and  detailed elemental
abundances.   We confirmed the slightly sub-solar metallicity of Be~39
([Fe/H]=$-$0.2)  and the very small scatter around the average abundances for
all elements measured (Fe, O, Na, Al, $\alpha$-, iron-peak, and heavy elements).
This is interesting also in the context of cluster and disc formation mechanisms
and disc chemical evolution.

The observations were optimised to look in particular for possible star-to-star 
variations in O and Na, which are indicative of the existence of multiple
populations in this  cluster, in analogy to what is is found for the higher
mass, older GCs.  No such scatter or anti-correlation was found; and we
conclude  that Be~39 is a normal, homogeneous, single-population cluster.

\begin{acknowledgements} This work was partially funded by the  PRIN INAF 2009
''Formation and Early  Evolution of Massive Star Clusters''.  This research has
made use of the package CataPack, for which we warmly thank Paolo Montegriffo,
of the WEBDA, of the SIMBAD database, operated at CDS, Strasbourg, France and of
NASA's Astrophysical Data System. C.S. acknowledges support from the US National
Science Foundation  through grant AST-1211585. \end{acknowledgements}

\end{document}